\documentclass[12pt]{iopart}

\usepackage{iopams}  
\usepackage{adjustbox}
\usepackage[graphicx]{realboxes}
\usepackage{epstopdf}
\usepackage{rotating}
\usepackage{adjustbox}
\usepackage{float}
\usepackage{latexsym}
\usepackage{amssymb}
\usepackage{array}
\usepackage{longtable}

\begin{document}

\title[MT Wormholes in $f(R,T)$ gravity]{Morris-Thorne Wormholes in the modified $f(R,T)$ gravity}
\author{A Chanda$^1$, S Dey$^2$, B C Paul$^3$ }
\address{Department of Physics, University of North Bengal, Siliguri, Dist.  Darjeeling 734 014, West Bengal, India}
\ead{$^1$anirbanchanda93@gmail.com}
\ead{$^2$sagardey231@gmail.com}
\ead{$^3$bcpaul@associates.iucaa.in}
\vspace{10pt}
\date{}

\vspace{10pt}

\begin{abstract}
Wormhole solutions obtained by Morris and Thorne (MT) in general relativity (GR) is investigated in a modified theory of gravity. In the gravitational action, we consider $f(R,T)$ which is a function of the Ricci scalar ($R$)  and the trace of the energy momentum tensor   ($T$).  In the framework of a modified gravity described by $f(R,T)=R+\alpha R^{2}+\lambda T$, where $\alpha$ and $\lambda$ are  constants,  MT wormhole solutions (WH) with normal matter are obtained for a relevant  shape function. The energy conditions are probed at the throat and away from the throat of the WH. It is found that the  coupling parameters, $\alpha $ and $\lambda$ in the gravitational action play an important role to decide the matter composition needed. It is found that for a given $\lambda$,  WH exists  in the presence of exotic matter at the throat when $\alpha <0$.   However, it is demonstrated that the WH exists without exotic matter when $\alpha >0$  in the modified gravity. Two different  shape functions  are analyzed for the  WH solutions that admits with or without exotic matter. It is noted that in a modified gravity MT WH permits even with normal matter which is not possible in GR. It is shown that a class of WH solutions exist with anisotropic fluid for $\lambda \neq - 8 \pi$.  However,   for flat asymptotic regions  with anisotropic fluids WH solutions cannot be realized for $\lambda= - 8 \pi$.  All the energy conditions are found consistent with the hybrid shape function indicating existence of WH even with normal matter for $\lambda \rightarrow 0$.\\
\vspace{2.0cm}
\end{abstract}
 
\maketitle


\section{Introduction}
\label{intro}
 Einstein's theory of General Relativity (GR)  permits traversable wormholes which are considered to describe topological passage through a hypothetical tunnel like bridges connecting two distant regions of a universe or two different distant universes. The term {\it wormhole}  was first introduced  in 1957 by Wheeler and Misner \cite{1,2}. Subsequently the  interesting features of the wormhole  led to a spurt in research activities in theoretical astrophysics \cite{3,4}.  The traversable wormhole allows  a shortcut passage between two distant regions of space-time. The observed features of the WH solutions are employed for constructing hypothetical time machines \cite{5,6}. Theoretically it is proposed that black holes and wormholes are interconvertible structures and a stationary wormhole might  be the final stage of a evaporating black hole \cite{7}. In  the literature \cite{8,9,10},  it  is  found that astrophysical accretion of ordinary matter could convert wormholes into black holes. It is known that WH exists in GR  at the cost of violation of energy conditions of the matter. The energy-momentum tensor of the matter supporting such geometries violates the null energy condition (NEC) near the throat region of the WH \cite{11,12,13} for its existence.  A fundamental ingredient in wormhole physics is the flaring out condition of the throat, which can be obtained in  GR with the violation of the NEC. Matter that violates the NEC is called exotic matter.  
In cosmology exotic matter described by dark energy or phantom fluid \cite{14,15},  is required to accommodate the late accelerating   phase of the universe.  The  phantom fluid is considered  \cite {16} to realize wormholes in GR.    The  exotic matter at the throat of the wormhole signifies that an observer who moves through the throat with a radial velocity approaching  the speed of light will observe presence of negative energy density leading to the violation of the energy conditions. However, it will be interesting if one can construct WH  without exotic matter. 

Recently, there is a growing interest to  modify the gravitational sector of GR to accommodate the observed universe. One of the simplest modification of the Einstein-Hilbert action is the $f(R)$- theory of gravity, where the curvature scalar ($R$)  in the gravitational action is replaced by $f(R)$ which is a polynomial function of  $R$. In  $f(R)=R+\alpha R^{2}$ theory, Starobinsky \cite{19} first obtained early inflationary universe solution  long before the efficacy of the inflation was known.  The higher order gravity is important to look for other aspects in astrophysics.
Recently, WH solutions are obtained in $f(R)$ theory of gravity \cite{20,21}, it is  shown that Starobinsky model requires exotic matter.
Using specific shape functions and a constant redshift function, Lobo and Oliveira \cite{22}  explored  criteria  that affect the wormhole structures. Thin shell like wormholes with charge is obtained in the framework of $f(R)$ gravity and their stability probed under perturbations  \cite{23}. Sahoo {\it et al} \cite{23a} obtained WH where NEC is violated in the framework of higher curvature and phantom field.
Recently Godani and Samanta  \cite{24} studied wormhole solutions for different shape functions in $f(R)$ gravity  and found that wormholes are filled with phantom fluid. However, in a ghost-free scalar tensor model of dark energy admitting phantom behaviour it is shown by  Bronnikov and Starobinsky \cite {25}  that realistic WH are not permitted even  in the presence of electric and magnetic fields.
 Both  Brans-Dicke theory and $f(R)$ gravity are considered  to obtain WH, it is claimed \cite{26}  that  no vacuum WH  exists in Brans-Dicke theory but WH exists in $f(R)$-gravity  if it satisfies an extremum where the effective gravitational constant  changes its sign.
Bronnikov {\it et al.} \cite{26a} shown a no-go theorem in GR for obtaining wormhole solutions, according to that it excludes the existence of wormholes with flat and/or AdS asymptotic regions on both sides of the throat if the source matter is isotropic. 
 
Recently the modified  $f(R,T)$ -gravity which is an arbitrary function of the Ricci scalar ($R$)  and of the trace of the energy momentum tensor ($T$) are considered widely for understanding the observed universe  \cite{27}. 
 In the modified gravity framework  the random requirement on $T$  with conceivable contributions from both non-minimal coupling and unambiguous $T$ terms in the gravitational action  may have rich structure in understanding the universe. Consequently  $f(R,T)$ theory of gravity is important   in understanding many different aspects considering different functional forms of the curvature scalar $R$ and trace of the energy- momentum tensor $T$
 \cite{28,29,30}. 
 Azizi \cite{31} obtained wormhole solutions with a  shape function and which however satisfies the null energy condition in the framework of $f(R,T)$ gravity. Zubair $\it{et. al.}$ \cite{32} considered three types of fluid in a static spherically symmetric wormhole under $f(R,T)$ gravity formalism and analysed the energy conditions. 
 The modified $f(R,T)$- gravity is considered to obtain WH in the presence of phantom fluid and found that WH violates the NEC in radial case, unlike in the tangential case \cite{33}.  
In f(R,T) modified theory of gravity existence of  spherically symmetric wormhole solutions are examined with   non-commutative geometry in terms of Gaussian and Lorentzian distributions of string theory \cite{34}. Stable wormhole solutions are found to exist which are used to estimate the deflection angle and found that it diverges at the wormhole throat.
 A hybrid form of shape function is considered \cite{35} to examine the energy conditions in the framework of $f(R,T)=R+ f(T)$ gravity. The radial null energy  and weak energy conditions are found to satisfy with the shape function in the absence of exotic matter.  WH solutions are  admitted with anisotropic and isotropic fluids under certain conditions.
Godani and Samanta \cite{36} considered another  functional form of $f(R,T)$ -gravity relevant for dark energy model in cosmology to obtain  wormhole solutions and found that wormhole solutions  are permitted without the requirement of non-exotic matter for a given type of shape function.
In this paper we consider a modified   $f(R,T)=f(R)+\lambda T$, where $\lambda$ is a constant and  $f(R)=R+\alpha R^{2} $ to obtain wormhole solutions. It may be pointed out that  $T=0$ was employed in black hole physics, matter density perturbations \cite{37,38,39,40} after its introduction  in cosmology \cite{19}. Compact astrophysical objects are also analysed  in Starobinsky model  \cite{41,42,43,44}. Recently Sahoo $\it{et. al.}$ \cite{45} studied WH solutions using a modified gravity given by  $f(R,T)= f(R) + \lambda T $  with a shape function which was considered in Ref. (\cite{46}) to obtain WH in GR.  The negative values of $\lambda$ was considered  in $f(R, T)= R + \alpha R^2 + \lambda T$-gravity to obtain wornmole. The  coupling of the  geometry and matter in the modified gravity is taken up in recent times to investigate various issues in cosmology and astrophysics.  It is motivated us to study wormhole solutions as in the limit $\lambda \rightarrow 0$ leads to $f(R)$ theory.  It is important to determine the role of $\lambda$ in obtaining  wormhole solutions with different shape functions for matter composition. 
Although the reliability of the isotropy in the fluid description has been experimentally verified in many contexts, there are many situations for which anisotropies may originate both at high and low energy densities which was reported in the context of  compact objects \cite{65}. The anisotropic distribution of fluids was used to study magnetized accretion disks around Kerr black holes \cite{66} (see Ref. \cite{67} for further studies on black holes/wormholes from anisotropic fluids). Shaikh \cite{68} constructed a wide class of wormholes in Eddington-inspired Born-Infeld gravity with a stress energy which does not violate the weak or null energy condition. The existence of anisotropic solutions usually involves strong gravity effects, and in some cases they are in the realm where  effects arises from a quantum theory of gravity  \cite{69}.  The wormholes will be probed in the presence of anisotropic fluid in the modified gravity.
 
The paper is organised as follows: In sec.2, the fundamentals of $f(R,T)$ gravity is presented. The general form of the field equation is obtained by varying the gravitational action with respect to the metric. In sec.3, we have considered the static spherically symmetric metric to describe the WH geometry and the necessary conditions which are to be satisfied  making use of two different shape functions.  The field equations are determined for the WH metric. In sec. 4, we have considered two possible WH shape functions with their features, and in sec. 5, the physical analysis has been carried out. The validity of energy conditions are studied by plotting graph. The results are summarised in sec. 6 followed by a brief discussion.


\section{The Gravitational action  in $f(R,T)$ modified theory and the field equations} 

The modified  gravitational action is given by \cite{25}
\begin{equation}
S=\frac{1}{16\pi} \int{d^{4}x\sqrt{-g}f(R,T)}+\int{d^{4}x\sqrt{-g} \; L_{m}}
\end{equation}
where $f(R,T)$ is a function of Ricci scalar ($R$ )and the trace of the energy-momentum tensor ($T=g^{\mu \nu} T_{\mu \nu}$) , $g$ is the determinant of the metric and $L_{m}$ is the matter Lagrangian density. 
Varying the action $S$ with respect to the metric $g_{\mu \nu}$ the field equation is obtained, which is given by \cite{25}
\[
R_{\mu \nu} \; f_{R}(R,T)-\frac{1}{2}f(R,T)g_{\mu \nu}+(g_{\mu \nu} \Box-\nabla_{\mu}\nabla_{\nu})f_{R}(R,T)
\]
\begin{equation}
=8\pi T_{\mu \nu}-f_{T}(R,T)\Theta_{\mu \nu}-f_{T}(R,T)T_{\mu \nu}
\end{equation}
where we denote $f_{R}(R,T)\equiv \frac{\partial f(R,T)}{\partial R}$, $f_{T}(R,T)\equiv \frac{\partial f(R,T)}{\partial T}$, $T_{\mu \nu}$ is the energy-momentum tensor and $\Theta_{\mu \nu}$ is defined as
\begin{equation}
\Theta_{\mu \nu}=g^{\alpha \beta}\frac{\delta T_{\alpha \beta}}{\delta g^{\mu \nu}}=-2T_{\mu \nu}+g_{\mu \nu}L_{m}-2g^{\alpha \beta} \frac{\partial^{2}L_{m}}{\partial g^{\mu \nu} \partial g^{\alpha \beta}}
\end{equation}
in the above we consider natural units, $i.e$, speed of light, $c=1$ and gravitational constant, $G=1$.\\
The energy momentum tensor for anisotropic fluid \cite{4} is given by
\begin{equation}
T_{\mu \nu}=(\rho+p_{t})u_{\mu}u_{\nu}-p_{t}g_{\mu \nu}+(p_{r}-p_{t})X_{\mu}X_{\nu},
\end{equation}
where $\rho$, $ p_{r}$ and $p_{t}$ represent the energy density, radial pressure and tangential pressure respectively. In the above $u_{\mu}$ and $X_{\mu}$ denotes the four-velocity vector and the radial unit four vector respectively, and  satisfy the relations $u^{\mu}u_{\nu}=1$ and $X^{\mu}X_{\nu}=-1$. Here we consider the matter Lagrangian density as follows :   $L_{m}=-P$, consequently one can  rewrite eq. (3)  as
\begin{equation}
\Theta_{\mu \nu}=-2T_{\mu \nu}-Pg_{\mu \nu}
\end{equation}
where $P=\frac{p_{r} + 2 \; p_{t}}{3}$, and the trace of the energy-momentum tensor is $T=\rho-3P$. It may be mentioned here that  following   choices for the matter Lagrangian  (i) $L_{m}=\rho$,  (ii) $L_{m}=-P$ and   (iii) $L_{m}=T$ are found in the literature. In geometry-matter coupling gravity theories namely,  $f(R,T)$ theory, an extra force acts along the orthogonal direction to the direction of four velocities for a (non-)geodesic motion.  The extra force depends on the choice of the matter Lagrangian which however found to vanishes when $L_{m}=-P$. The effective Einstein field equation corresponding to the eq. (2) can be rewritten as
\begin{equation}
R_{\mu \nu}-\frac{1}{2} \; g_{\mu \nu} \; R = T^{eff}_{\mu \nu}
\end{equation}
where  the effective stress-energy tensor is 
\[
T_{\mu \nu}^{eff}=\frac{1}{f_{R}(R,T)}\Big[(8\pi+f_{T}(R,T))T_{\mu \nu}+Pg_{\mu \nu}f_{R}(R,T)\Big]
\]
\[
\; \; \; \; + \frac{1}{f_{R}(R,T)}\Big[\frac{1}{2}[f(R,T)-Rf_{R}(R,T)]g_{\mu \nu}\Big]
\]
\begin{equation}
\; \; \; \; \; -  \frac{1}{f_{R}(R,T)}\Big[(g_{\mu \nu}\Box-\nabla_{\mu}\nabla_{\nu})f_{R}(R,T)\Big].
\end{equation}
The effective stress energy tensor ($ T^{eff}_{\mu\nu}$) is determined by  the matter stress-energy tensor $T_{\mu\nu}$ and the curvature quantities originating  from the  $f(R,T)$ modified theory of gravity. \\
The covariant derivative of the stress-energy tensor is given by 
\begin{equation}
\nabla^{\mu}T_{\mu \nu}=\frac{f_{T} [ (T_{\mu \nu}+\Theta_{\mu \nu})\nabla^{\mu}  ln f_{T}
+\nabla^{\mu}\Theta_{\mu \nu}-\frac{1}{2} \; g_{\mu \nu}\nabla^{\mu}T ]}{8\pi-f_{T}}
\end{equation}
where  $f_T=\frac{\partial f(R,T)}{\partial T}$. It may be pointed out here that the covariant divergence of the stress-energy tensor $T_{\mu \nu}$ in $f(R,T)$ theory of gravity is not conserved unlike GR or $f(R)$ theories of gravity.
However, in the $f(R, T)= R + \alpha R^2 + \lambda T$ theory of gravity, the covariant derivative of effective energy momentum tensor $ T^{eff}_{\mu\nu}$ is zero, {\it i.e.}, 
\begin{equation}
\nabla^{\mu}T^{eff}_{\mu\nu} = 0.
\end{equation}
The geometry matter coupling in $f(R,T)$ theory leads to a non-vanishing of the four divergence of the energy momentum tensor hence it is the effective energy momentum tensor that represents the conservation equation. Thermodynamical interpretation of this geometry matter coupling has been studied by Harko\cite{70}  and it is 
pointed out  that the non-conservation of the matter energy momentum tensor is related with irreversible matter creation process. It is also shown that the model parameters decides the creation pressure and the rate  of particle production.  It is also pointed out that the types of particles created can not be predicted with certainty. Using quantum analogy,  it was predicted that most of the particles created in such geometry matter coupling may be due to some scalar particles or bosons which   in the cosmological scale may contribute to the dark matter content in the universe.

\section{Wormholes in $f(R,T)$ Gravity }

In this section static spherically symmetric metric is considered for describing  wormholes which is given by \cite{4,45}
\begin{equation}
ds^{2}=e^{2\phi(r)}dt^{2}- e^{2 \gamma (r)} dr^{2}-r^{2}(d\theta^{2}+sin^{2}\theta d\phi^{2})
\end{equation}
where $\phi(r)$  denote the redshift function and for wormhole geometry we consider
$e^{2 \gamma(r)} = \frac{1}{1-\frac{b(r)}{r}}$, $b(r) $  is the
shape function. The radial coordinate $r$ in this case increases from a minimum value $r_{0}$ to $\infty$, $r_{0}$ being the WH throat radius.  The WH metric must satisfy the following  conditions as mentioned below: \\

$\bullet$ The range of  radial coordinate $r$  is $r_{0} \le r \le \infty$, with $r_{0}$ being the throat radius.\\
\\

$\bullet$ The shape function $(b(r))$ satisfies the condition $b(r_{0})=r_{0}$, at the throat and away from the throat $\it{i.e.}$ for $r>r_{0}$ it must satisfy the constraint condition
\begin{equation}
\hspace{2.5 cm} 1-\frac{b(r)}{r} > 0.
\end{equation}

$\bullet$ For a physical  flaring out condition, it must be satisfied by the shape function $b(r)$ which at the throat of a WH solution becomes $\it{i.e.}$ $b'(r_{0})<1$.\\
\\

$\bullet$ For an asymptotic flatness of the spacetime geometry one obtains
\begin{equation}
\; \hspace{2.0 cm} \frac{b(r)}{r} \rightarrow 0 \ \ \ as \ \ \ |r| \rightarrow \infty
\end{equation}

$\bullet$ At the throat $r_{0}$, the redshift function $\phi(r)$ must be a  finite non-vanishing function.\\
\\
For simplicity we assume here that the redshift function is a  constant  ($\phi \rightarrow 0$) accommodating the asymptotic de Sitter or anti-de Sitter solution of the WH metric \cite{44}.
The effective field equations for the modified gravitational theory $f(R,T)=R+\alpha R^{2}+\lambda T$ are obtained for the wormhole metric given by eq. (10)  are
\begin{equation}
\frac{b'}{r^{2}}=\frac{\Big[(8\pi+\frac{3\lambda}{2})\rho-\frac{\lambda(p_{r}+2p_{t})}{6}-\frac{2\alpha b'^{2}}{r^{4}}\Big]}{2\alpha R+1},
\end{equation}
\begin{equation}
\frac{b}{r^{3}}=\frac{\Big[-(8\pi+\frac{7\lambda}{6})p_{r}+\frac{\lambda}{2}(\rho-\frac{2p_{t}}{3})-\frac{2\alpha b'^{2}}{r^{4}}\Big]}{2\alpha R+1},
\end{equation}
\begin{equation}
\frac{b'r-b}{2r^{3}}=\frac{\Big[-(8\pi+\frac{4\lambda}{3})p_{t}+\frac{\lambda}{2}(\rho-\frac{p_{r}}{3})-\frac{2\alpha b'^{2}}{r^{4}}\Big]}{2\alpha R+1}.
\end{equation}
The Ricci scalar for the WH metric is given by
\begin{equation}
R=\frac{2b'}{r^{2}}.
\end{equation}
The components of the effective stress energy tensor ($T_{\mu\nu}^{eff}$) are determined as
\begin{equation}
\rho_{eff}=\frac{\Big[(8\pi+\frac{3\lambda}{2})\rho-\frac{\lambda(p_{r}+2p_{t})}{6}-\frac{2\alpha b'^{2}}{r^{4}}\Big]}{2\alpha R+1},
\end{equation}
\begin{equation}
p_{r(eff)}=\frac{\Big[-(8\pi+\frac{7\lambda}{6})p_{r}+\frac{\lambda}{2}(\rho-\frac{2p_{t}}{3})-\frac{2\alpha b'^{2}}{r^{4}}\Big]}{2\alpha R+1},
\end{equation}
\begin{equation}
p_{t(eff)}=\frac{\Big[-(8\pi+\frac{4\lambda}{3})p_{t}+\frac{\lambda}{2}(\rho-\frac{p_{r}}{3})-\frac{2\alpha b'^{2}}{r^{4}}\Big]}{2\alpha R+1}.
\end{equation}
In GR, a fundamental point in wormhole physics is the energy condition violations, it needs to be investigated the energy conditions  in modified theories of gravity  as  the gravitational field equations differ
from the classical relativistic Einstein equations.
In modified gravity we have generalized NEC, $i.e.$, $T_{\mu\nu}^{eff}K^{\mu}K^{\nu}<0$, where $K^{\mu}$ is a  null vector \cite{63}.  The generalized NEC reduces to that in GR for  $\lambda =0$ and $\alpha=0$. In the $f(R, T)$-theory, we get  a different picture where in principle  the matter stress energy tensor satisfies the standard NEC, $i.e.$, $T_{\mu\nu}K^{\mu}K^{\nu}\ge0$, while the respective generalised NEC is violated in order to ensure the flaring out condition. Using the field equations the generalized NEC is obtained as follows
\begin{equation}
\rho_{eff}+p_{r(eff)}=\frac{b+b'r}{r^3}<0.
\end{equation}
The effective anisotropy can be expressed as
\[
p_{t(eff)}-p_{r(eff)} =\frac{8 \pi + \lambda }{1+2 \alpha R} \;  \Delta 
\]
where $\Delta= p_r-p_t$ is the measure of anisotropy.
It is evident that the effective anisotropic pressure  vanishes if (i) $p_r=p_t$  or (ii) $\lambda = - 8 \pi$.  For static 
spherically symmetric space-times in  $f(R, T)$-gravity,  the no go theorem of general relativity that 
it excludes the existence of wormholes with flat and/or AdS asymptotic regions on both sides of the throat with isotropic source of matter  \cite{26a} for $\phi+\gamma=0$ in eq. (10) is also true. In addition to that $f(R,T)$-theory add one more  criteria that no go theorem is also valid with anisotropic fluid source when $\lambda= - 8 \pi$.  In the next section we obtain wormhole solutions  for  $\lambda \neq -8 \pi$ with anisotropic fluid.

 Using eqs. (13)-(16),  energy density, radial pressure and tangential pressure can be rewritten as 
 \begin{equation}
\rho=\frac{b'\Big[\lambda(2r^{2}+11\alpha b')+12\pi (r^{2}+6\alpha b')\Big]}{3(\lambda+4\pi)(\lambda+8\pi)r^{4}}
\end{equation}
\begin{equation}
p_{r}=\frac{X}{3(\lambda+4\pi)(\lambda+8\pi)r^{5}}
\end{equation}
\begin{equation}
p_{t}=-\frac{Y}{6(\lambda+4\pi)(\lambda+8\pi)r^{5}}
\end{equation}
where, $X=-3b(r^{2}+4b'\alpha)(4\pi+\lambda)+b'r(r^{2}\lambda+b'\alpha(-24\pi+\lambda))$ and $Y=3b(r^{2}+4b'\alpha)(4\pi+\lambda)-b'r(12\pi(r^{2}+8b'\alpha)+(r^{2}+10b'\alpha)\lambda )$.\\
The EoS for tangential pressure is
\begin{equation}
p_{t}=\omega \; \rho
\end{equation}
where, the equation of state parameter $\omega$  is a function of $r$.
We consider here anisotropic fluid distribution where the radial and tangential pressures are different and the corresponding EoS  is 
\begin{equation}
p_{r}=\omega_{1} \; p_{t}
\end{equation}
where $\omega_{1}$ is also a function of $r$. It may be mentioned here that  wormhole solutions are obtained considering   a hyperbolic function for $\omega_{1}$  \cite{33}.  But it is not necessary to assume a  functional form of the EoS parameter. As the field equations are highly non-linear, and the EoS state parameters  varies with $r$ in addition to its  dependence on the coupling parameters $\alpha$ and $\lambda$ of the gravitational action, we study numerically. 
There are three equations and four unknowns namely, $\rho$, $p_r$, $p_t$ and $b(r)$, therefore, we choose the shape functions for wormholes.  For a given shape function $b(r)$, we determine the matter necessary for wormhole solution.
The effective equation of state  parameters represented in eqs. (24) and (25)  are given by
\begin{figure}[H]
\begin{equation}
\resizebox{0.5\textwidth}{!}{$\omega=\frac{3b(r^{2}+4b'\alpha)(4\pi+\lambda)-b'r[12\pi(r^{2}+8b'\alpha)+(r^{2}+10b'\alpha)\lambda]}{2b'r[12\pi(r^{2}+6b'\alpha)+(2r^{2}+11b'\alpha)\lambda]}$},
\end{equation}
\begin{equation}
\resizebox{0.5\textwidth}{!}{$\omega_{1}=-\frac{2[3b(r^{2}+4b'\alpha)(4\pi+\lambda)+b'r[b'\alpha(24\pi-\lambda)-r^{2}\lambda]]}{3b(r^{2}+4b'\alpha)(4\pi+\lambda)-b'r[12\pi(r^{2}+8b'\alpha)+(r^{2}+10b'\alpha)\lambda]}$}.
\end{equation}
\end{figure}
For a given shape function the variation of the EoS parameters  are determined with $\alpha$ and $\lambda$ of the gravitational action.

\section{Wormhole Models}

 In this section  wormhole solutions are investigated  with the following two different shape functions : \cite{35,49} (I).  $b(r)=r_{0} \; e^{1-\frac{r}{r_{0}}}$ and (II). $b(r)=r_{0}^{2} \; \frac{e^{r_{0}-r}}{r}$. {\it The wormhole models considered here are asymptotically flat as for $r\rightarrow \infty$, $\left(1-\frac{b(r)}{r} \right)\rightarrow 1$.}\\

{\bf Case I :} {\bf {Shape Function : $b(r)=r_{0}\, e^{1-\frac{r}{r_{0}}}$}.} \\

The shape function  satisfies all the criterion for accommodating wormholes as mentioned in section 3 \cite{49}. The throat radius of the WH  is at $r_{0}=0.5$. Using eqs. (21)-(23) and (26), (27),  we get  $\rho$, $p_{r}$ and $p_{t}$  as  follows
\begin{equation}
\rho=\frac{e^{1-\frac{2r}{r_{0}}}(-2e^{\frac{r}{r_{0}}}r^{2}(6\pi+\lambda)+e\alpha(72\pi+11\lambda))}{3r^{4}(32\pi^{2}+12\pi\lambda+\lambda^{2})},
\end{equation}
\begin{equation}
\resizebox{0.8\textwidth}{!}{$p_{r}=-\frac{e^{1-\frac{2r}{r_{0}}}(e^{\frac{r}{r_{0}}}r^{2}(12\pi r_{0}+(r+3r_{0})\lambda)+e\alpha(24\pi(r-2r_{0})-(r+12r_{0})\lambda))}{3r^{5}(32\pi^{2}+12\pi\lambda+\lambda^{2})}$},
\end{equation}
\begin{equation}
\resizebox{0.8\textwidth}{!}{$p_{t}=\frac{e^{1-\frac{2r}{r_{0}}}(e^{\frac{r}{r_{0}}}r^{2}(12\pi(r+ r_{0})+(r+3r_{0})\lambda)-2e\alpha(24\pi(2r+r_{0})+(5r+6r_{0})\lambda))}{6r^{5}(32\pi^{2}+12\pi\lambda+\lambda^{2})}$}.
\end{equation}
The energy-density, radial pressure and tangential pressure  are functions of the coupling parameters $\lambda$ and $\alpha$ of the gravitational action. 
The energy conditions can be  studied for a wide range of values of the parameters wormholes. 
\\

{\bf Case II :} { \bf {Shape Function II $b(r)=r_{0}^{2}\, \frac{e^{r_{0}-r}}{r}$}.} \\

In this case the WH solutions are obtained for the shape function $b(r)=r_{0}^{2}\frac{e^{r_{0}-r}}{r}$ \cite{35}. In this case we get the throat radius  at  $r_{0}=1$.  The energy density ($\rho$), radial pressure and tangential pressure are given by 
\begin{figure}[H]
\begin{equation}
\resizebox{0.8\textwidth}{!}{$\rho=\frac{e^{-2r+r_{0}}(1+r)r_{0}^{2}(-2e^{r}r^{4}(6\pi+\lambda)+e^{r_{0}}(1+r)r_{0}^{2}\alpha(72\pi+11\lambda))}{3r^{8}(32\pi^{2}+12\pi\lambda+\lambda^{2})}$},
\end{equation}
\begin{equation}
\resizebox{0.8\textwidth}{!}{$p_{r}=-\frac{e^{-2r+r_{0}}r_{0}^{2}(e^{r}r^{4}(12\pi+(4+r)\lambda)+e^{r_{0}}(1+r)r_{0}^{2}\alpha(24\pi(-1+r)-(13+r)\lambda))}{3r^{8}(32\pi^{2}+12\pi\lambda+\lambda^{2})}$},
\end{equation}
\begin{equation}
\resizebox{0.8\textwidth}{!}{$p_{t}=-\frac{e^{-2r+r_{0}}r_{0}^{2}(-e^{r}r^{4}(12\pi(2+r)+(4+r)\lambda)+2e^{r_{0}}(1+r)r_{0}^{2}\alpha(24\pi(3+2r)+(11+5r)\lambda))}{6r^{8}(32\pi^{2}+12\pi\lambda+\lambda^{2})}$}.
\end{equation}
\end{figure}

\section{Physical Analysis}

The field equations are complex  functional form, therefore, the physical features of the WH geometry will be studied numerically in this section. The radial variation of energy conditions namely, dominant energy condition (DEC), weak energy condition (WEC),  null energy condition (NEC) and the strong energy condition (SEC) are plotted to examine the validity in both the cases.
 We probe the admissibility of matter from  the validity of the following  energy conditions  in the entire geometry for wormholes : \\
 
1. DEC: $\rho\ge|p_{i}|$,\\ 

2. WEC : $\rho\ge0$; $\rho+p_{i}>0$;\\

3. NEC: $\rho+p_{i}\ge0$;\\

4. SEC: $\rho+p_{r}+2p_{t}\ge 0$;\\

where $i=r,t.$\\

 The energy conditions are important to look for wormhole solutions  in the modified gravity with  normal or exotic matter  in the universe. We try to solve the problem of exotic matter by investigating wormhole in modified theories
of gravity. For  this  two different models with different shape functions are considered in the next section.

\subsection{\bf Model I:}

The  energy density ($\rho$), the radial pressure ($p_{r}$) and  the tangential pressure ($p_{t}$) for the shape function  $b(r)=r_{0}e^{1-\frac{r}{r_{0}}}$  given in Case I, are given in eqs. (28)-(30). Using the equations  the Energy conditions WEC, NEC, DEC and SEC are analyzed.\\

The positivity of the energy density ($\rho$) at the throat gives rise to the following constraints between $\alpha $ and $\lambda$: (i) for $\alpha >  \left(\frac{2(\lambda + 6 \pi)}{(11 \lambda + 72 \pi)} \right)$ , (ii)  $\alpha > \frac{1}{6}$ with $\lambda=0$, (iii) Both $\alpha$ and $\lambda$ are negative satisfying the inequalities $|\alpha| <  \left(\frac{2(|\lambda| -6 \pi)}{(72 \pi - 11 |\lambda|)} \right)$ with $6 \pi < | \lambda| < \frac{72 \pi}{11}$, (iv) For a positive $\alpha$, the following  limiting values are found $\alpha < \left(\frac{2(|\lambda | -6 \pi)}{ (11 |\lambda| -72 \pi)} \right)$ and $\frac{72 \pi }{11} <|\lambda | <6 \pi$.\\

\begin{figure}[H]
\begin{center}
\includegraphics[scale=0.55]{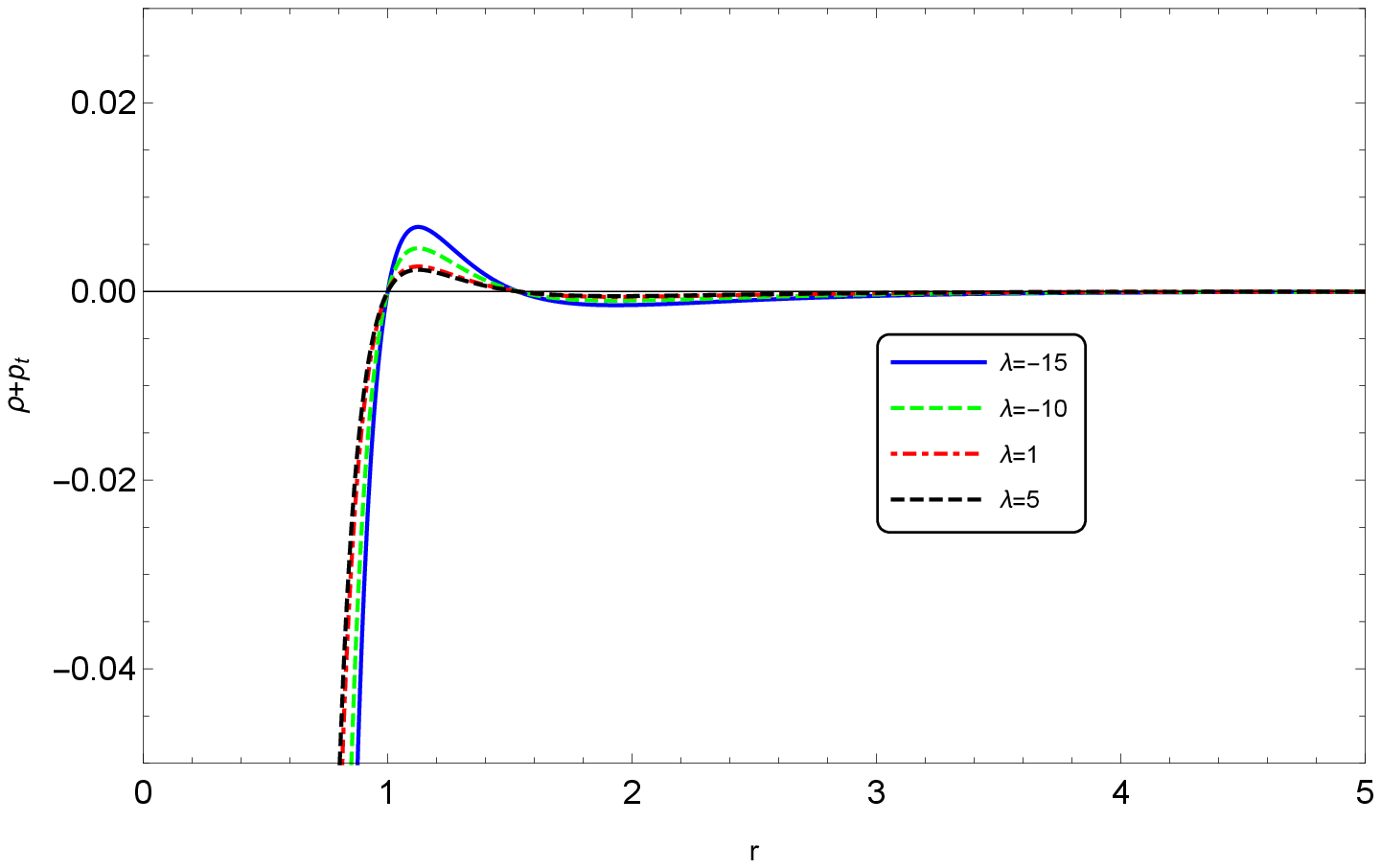}
\caption{$\rho+p_{t}$ with $\alpha=1$ and different $\lambda$ for shape function I.}
\label{fig: 1}
\end{center}
\end{figure}

\begin{figure}[H]
\begin{center}
\includegraphics[scale=0.55]{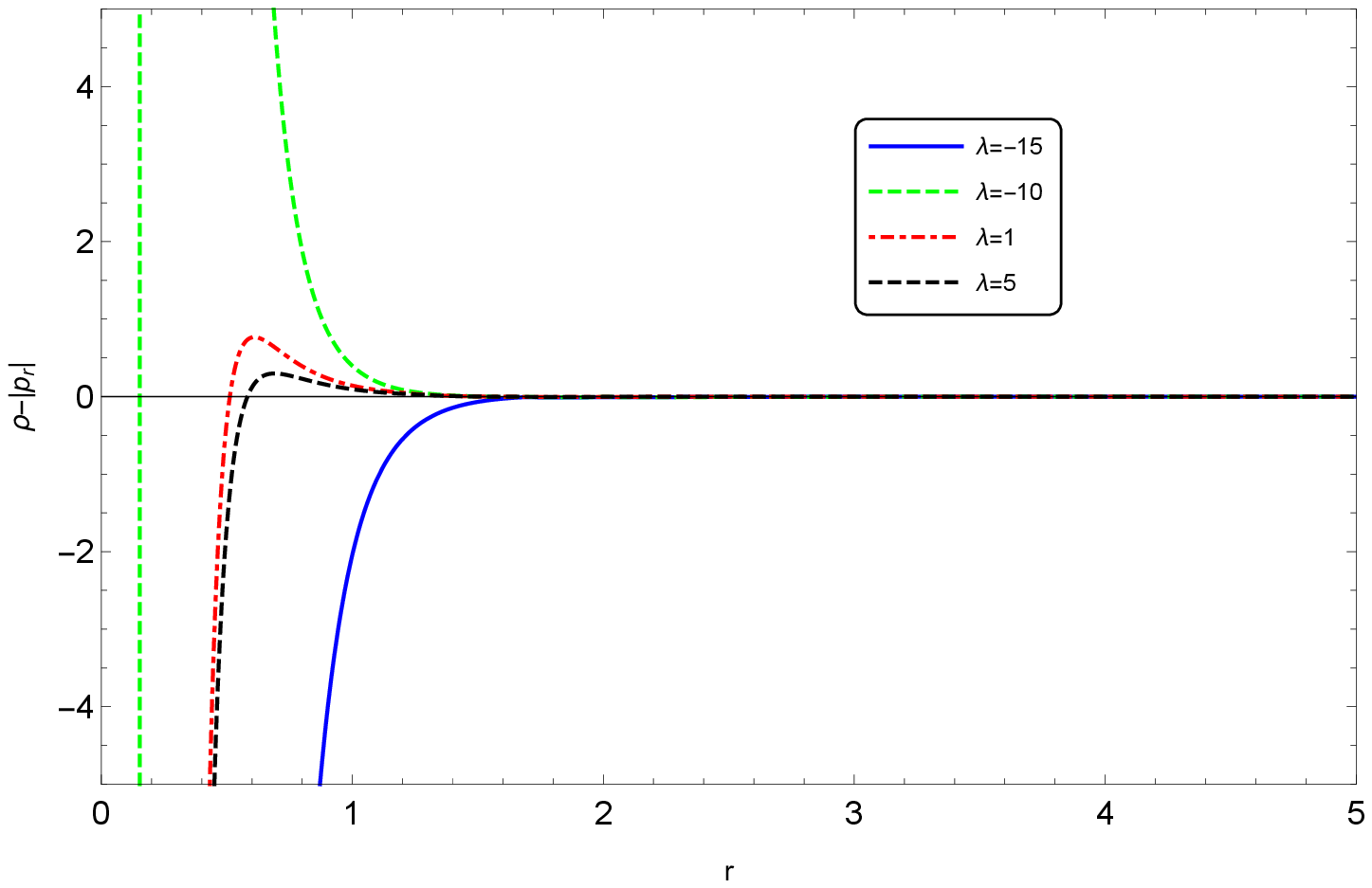}
\caption{ $\rho - |p_{r}|$ with $\alpha=1$ and different $\lambda$ for shape function I.}
\label{fig: 2}
\end{center}
\end{figure}

\begin{figure}[H]
\begin{center}
\includegraphics[scale=0.55]{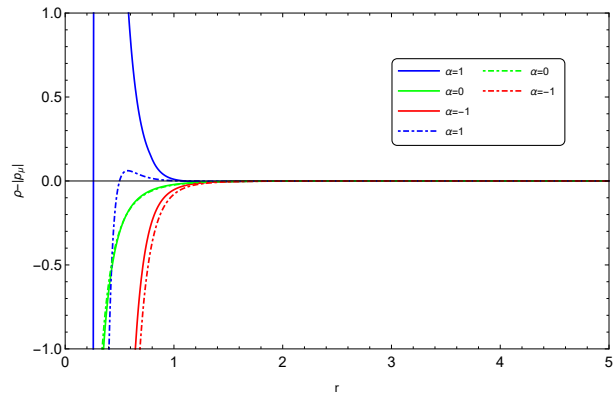}
\caption{Radial variation of  $\rho + p_{r}$ (Solid line) and $\rho + p_{t}$ (Dot Dashed lines) with $\lambda=1$ and different $\alpha$ for shape function I.}
\label{fig: 3}
\end{center}
\end{figure}

For  simplicity, the radial variation of the NEC is checked in Fig. (1) for $\alpha =1$ with different $\lambda$. The plot shows that the NEC, $\rho+p_{t}\ge 0$ is not satisfying near the throat implying  violation of the NEC.  It is also found that away from the throat NEC is valid and then again it fails to satisfy but at asymptotic region it obeys again. The  radial variation of $\rho-|p_{r}| $ for different $\lambda$ is plotted with $\alpha =1$  in Fig.(2). 
The radial variation of NEC for different $\alpha$  is checked in Fig. (3), it is found that NEC is violated at the throat which obeys away from  the throat.
In Fig. (4) the radial variation of energy density is plotted for three different cases of $\alpha$, it is found that $\rho > 0$ only for positive $\alpha$. 
In Fig. (5), radial variation of DEC is checked for different $\alpha$, it is found  that DEC  is violated. Thus the shape function I permits wormholes with exotic matter at the throat in the $f(R, T)$-modified gravity.

\begin{figure}[H]
\begin{center}
\includegraphics[scale=0.55]{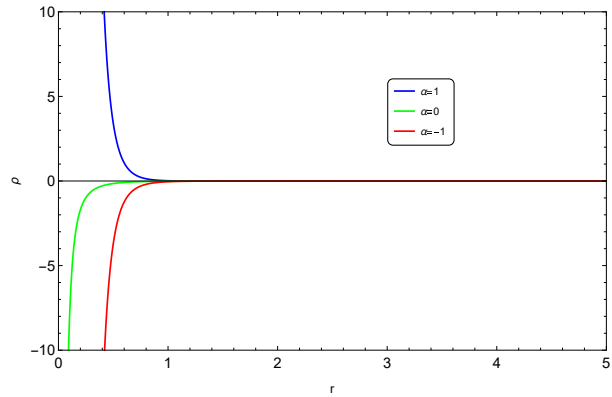}
\caption{Radial variation of Energy Density ($\rho$ ) with $\lambda=1$ with different $\alpha$ for shape function I.}
\label{fig: 4}
\end{center}
\end{figure}

\subsection{\bf Model II}
We consider here hybrid shape function given by $b(r)=r_{0}^{2}\frac{e^{r_{0}-r}}{r}$ for wormhole solutions.  The energy density ($\rho$) at the throat is positive when (i) for positive $\alpha >  \left(\frac{2(\lambda + 6 \pi)}{11 \lambda + 72 \pi} \right)$, (ii) $\alpha > \frac{1}{6}$ with $\lambda =0$,
(iii)  Both $\alpha$ and $\lambda$ are negative satisfying the inequalities $| \alpha | <  \left(\frac{2(|\lambda| -6 \pi)}{ 72 \pi - 11 |\lambda|} \right)$ with $6 \pi < | \lambda| < \frac{72 \pi}{11}$, 
 (iv)  $\alpha$  positive satisfying the inequalities $\alpha > \left(\frac{2 ( 6 \pi - |\lambda |) }{72 \pi - 11 | \lambda | } \right)$ and $ \frac{72 \pi }{11} <| \lambda |<6\pi $.

\begin{figure}[H]
\begin{center}
\includegraphics[scale=0.55]{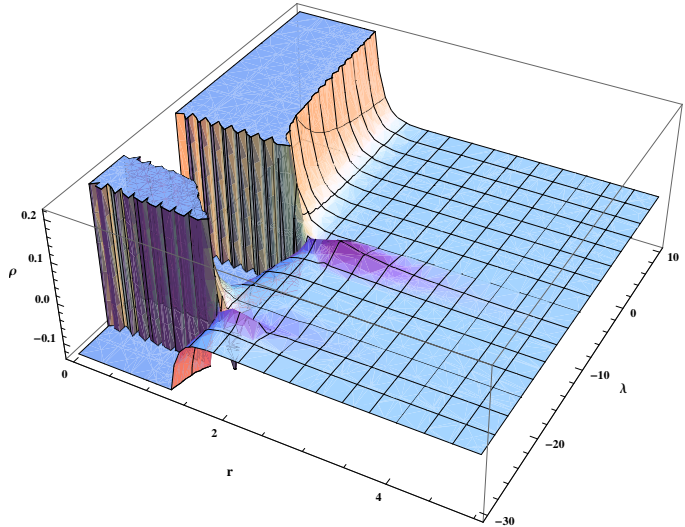}
\caption{Radial variation of energy density for $\alpha=1$ and different $\lambda$ for shape function II.}
\label{fig: 6}
\end{center}
\end{figure}

\begin{figure}[H]
\begin{center}
\includegraphics[scale=0.55]{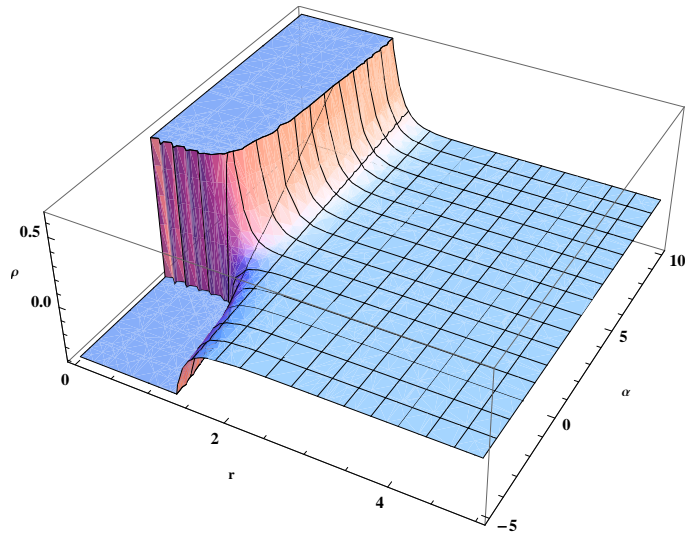}
\caption{Radial variation of Energy Density $\rho$  with different  $\alpha$ when $\lambda=1$ for shape function II.}
\label{fig: 7}
\end{center}
\end{figure}

\begin{figure}[H]
\begin{center}
\includegraphics[scale=0.55]{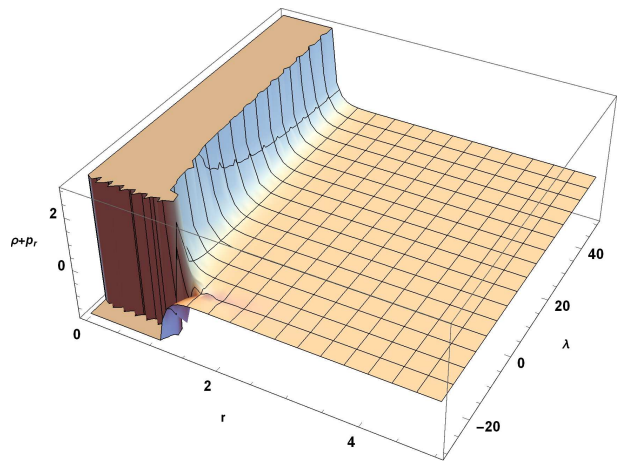}
\caption{Radial variation of NEC, $\rho+p_{r}\ge0$ with $\alpha=1$ and different $\lambda$ values for shape function II.}
\label{fig: 8}
\end{center}
\end{figure}

\begin{figure}[H]
\begin{center}
\includegraphics[scale=0.55]{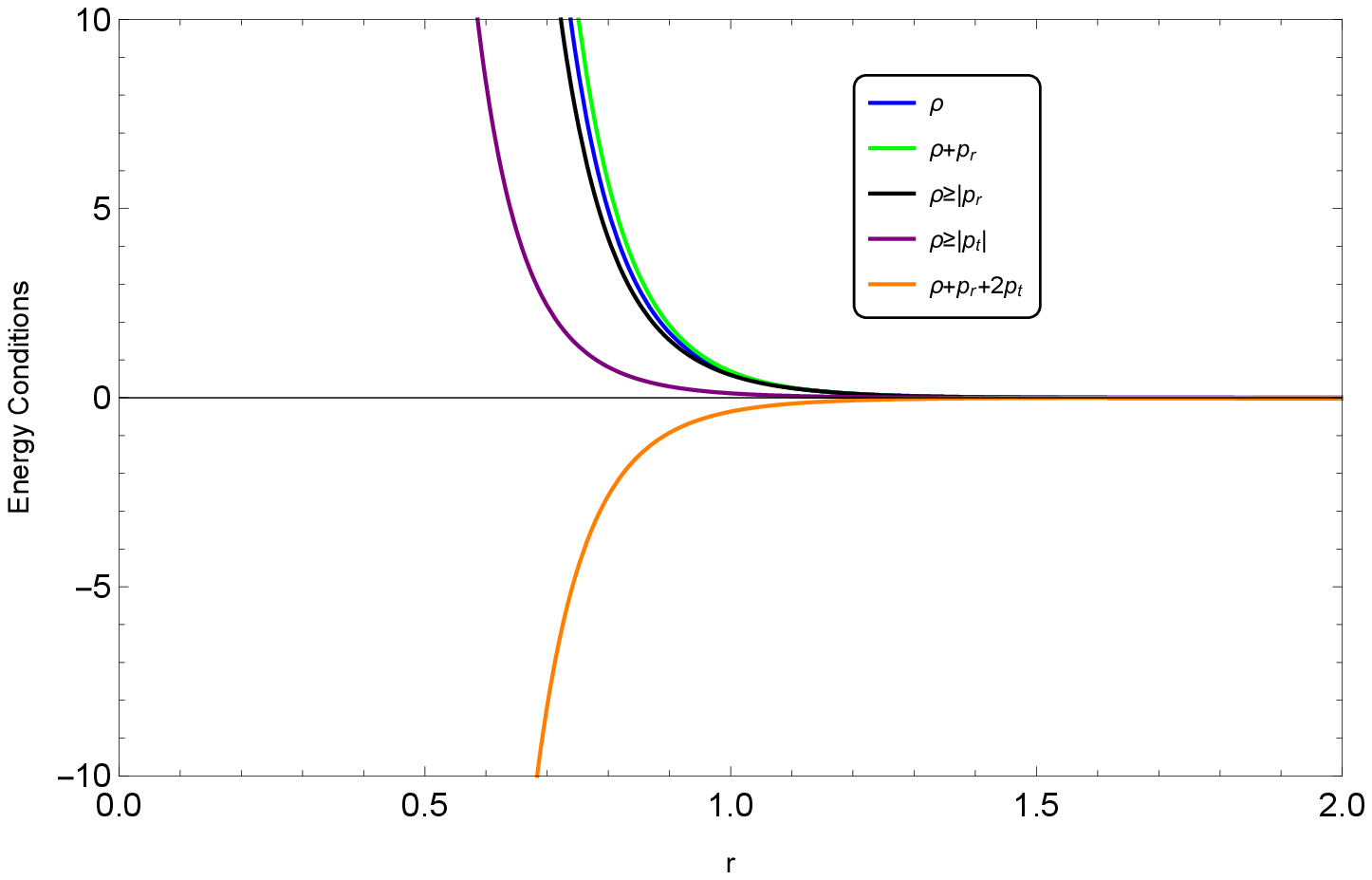}
\caption{Radial variations of the Energy Conditions for $\lambda=5$ and $\alpha=1$ for shape function II. }
\label{fig: 9}
\end{center}
\end{figure}

\begin{figure}[H]
\begin{center}
\includegraphics[scale=0.55]{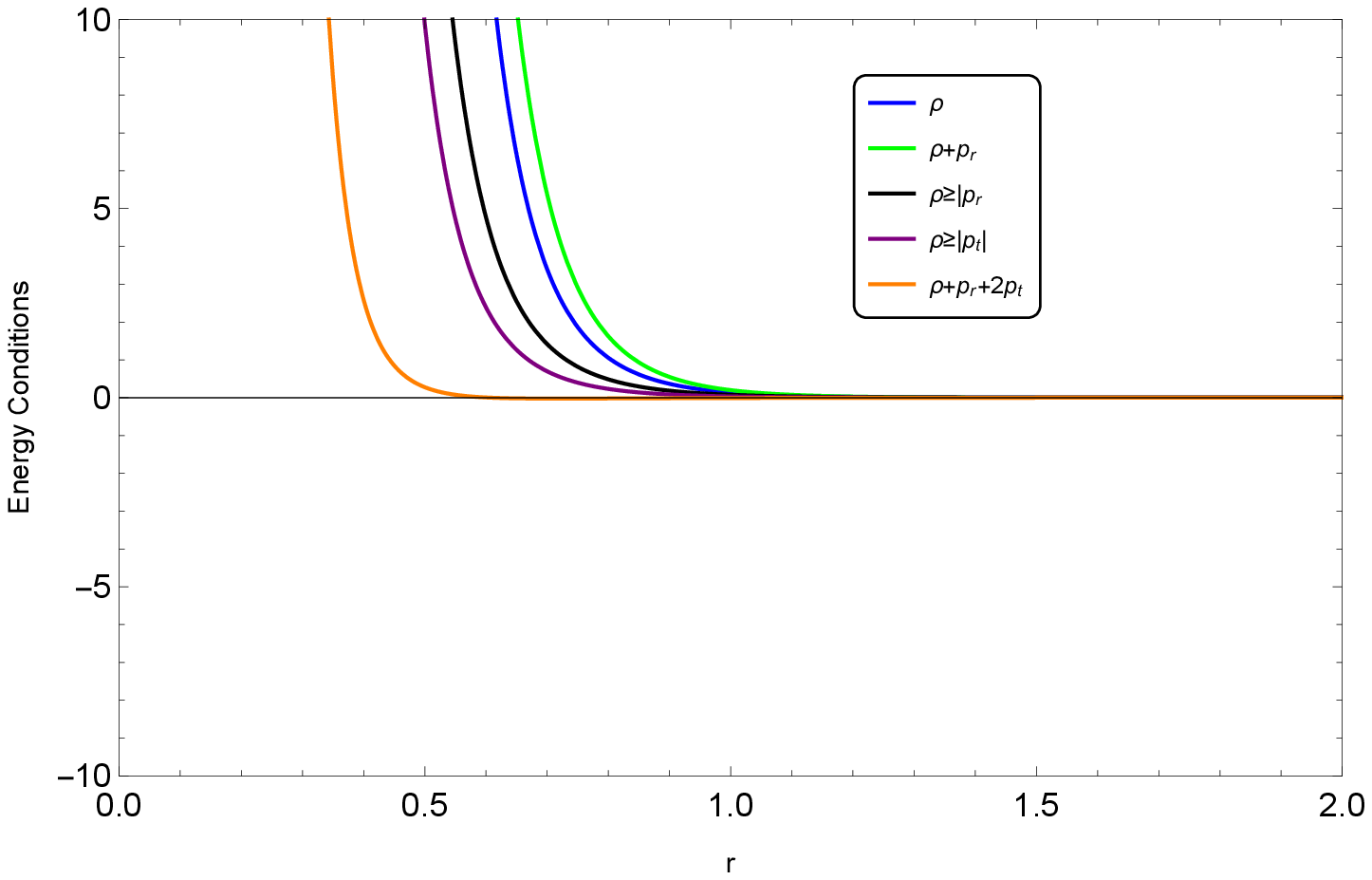}
\caption{Radial variations of the Energy Conditions for $\lambda=80$ and $\alpha=1$ for shape function II. }
\label{fig: 10}
\end{center}
\end{figure}

\begin{figure}[H]
\begin{center}
\includegraphics[scale=0.5]{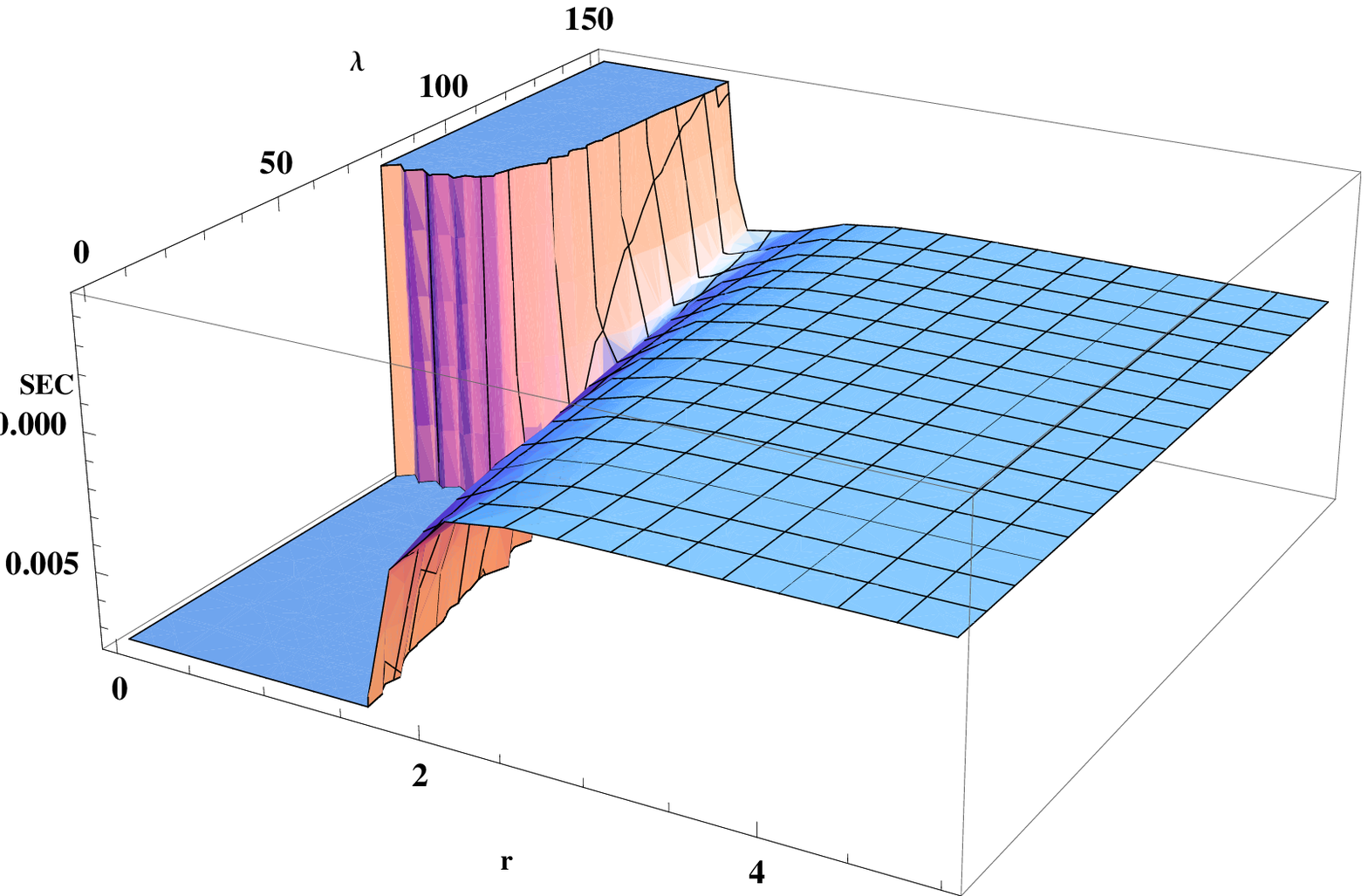}
\caption{Radial variation of SEC for different $\lambda$ when  $\alpha =1$  for shape function II.}
\label{fig: 11}
\end{center}
\end{figure}

\begin{figure}[H]
\begin{center}
\includegraphics[scale=0.55]{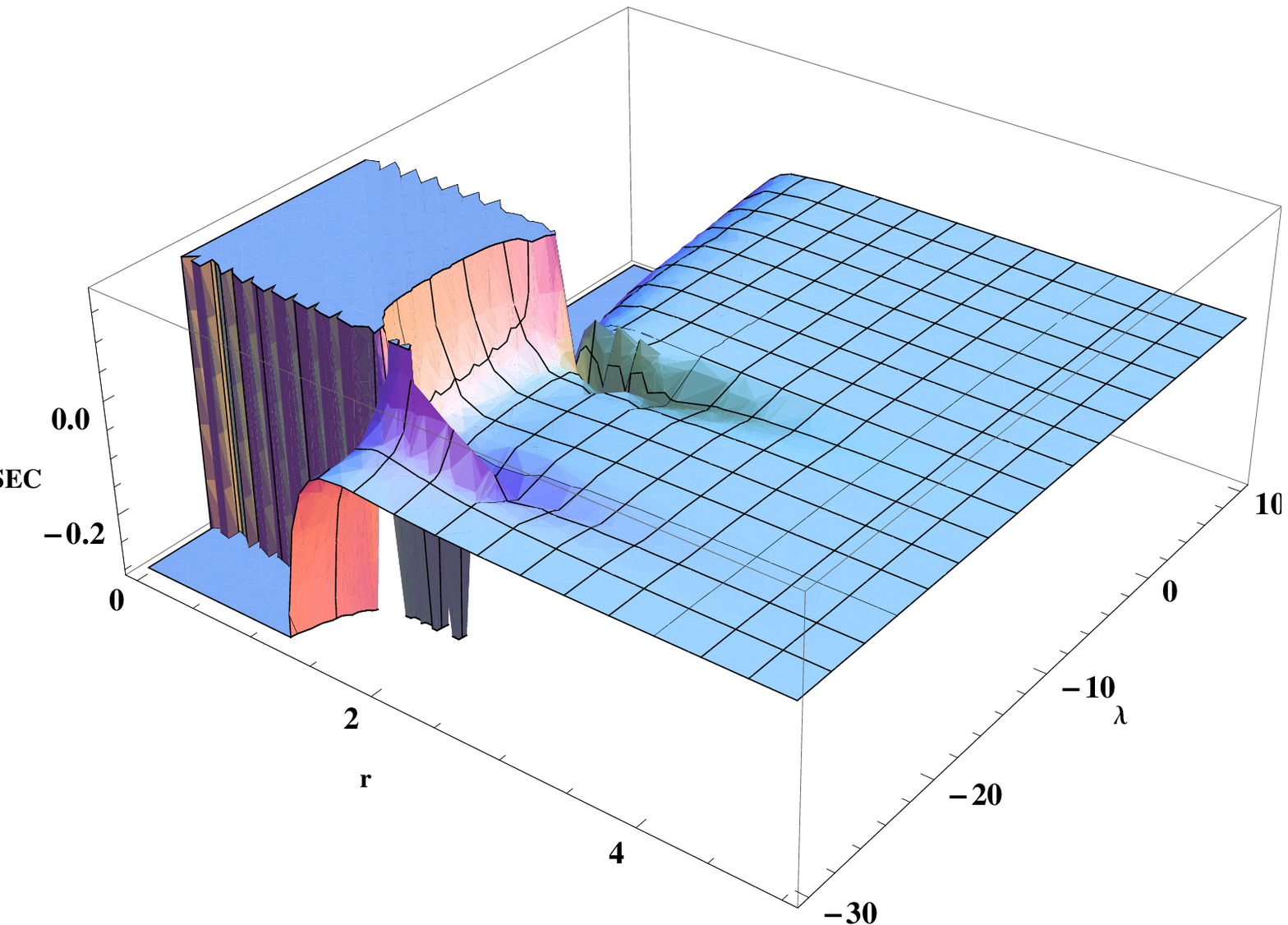}
\caption{Radial variations of SEC for different  $\lambda$ with $\alpha=2$ for shape function II. }
\label{fig: 12}
\end{center}
\end{figure}

\begin{figure}[H]
\begin{center}
\includegraphics[scale=0.55]{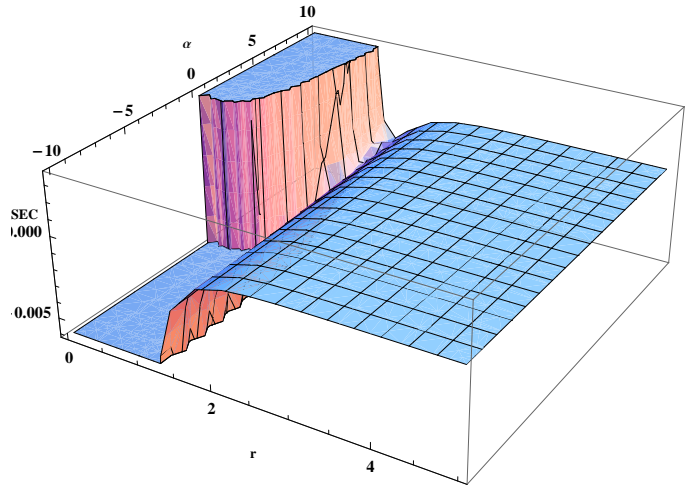}
\caption{SEC with different $\alpha$ for $\lambda=80$ for shape function II.}
\label{fig: 13}
\end{center}
\end{figure}

The 3D plot of the energy density  for  a range of $\lambda$ values taking  $\alpha =1$ in Fig. (5) shows that there is a range of $\lambda$ values for which $\rho$ is positive. For $\alpha=1$, $\rho$ is positive for $\lambda > -10$. There is a small range of negative $\lambda$ values for which it is positive thereafter it becomes negative again. 
It is evident from 3D plot in Fig. (7) that for $\lambda=1$, energy density is always positive for $\alpha > 0$ but energy density becomes negative  for $\alpha \le0$. Thus wormhole models will be studied in this case with positive $\alpha$ for different values of  $\lambda$. We note that wormholes solutions exist with normal matter when   $\alpha >0$ and $\lambda=0$, this result is different from that obtained in Ref. (\cite{36}).

A 3D plot of radial variation of NEC with different values of $\lambda$ for $\alpha=1$ in Fig. (8) shows that there exists a lower bound  $\lambda > -8 \pi$ for which  NEC is found to satisfy throughout in wormhole geometry.
 We  consider $\alpha=1$ and $\lambda=5$ to plot all the energy conditions in Fig. (9), the energy conditions except the SEC are found to satisfy.   However, for a large value say, $\lambda=80$, we plot all the energy conditions in Fig (10), it is clear from the plot that all the energy conditions are satisfied  accommodating wormholes.   This is interesting as traversable wormhole are permitted with normal matter in the modified gravity.  A 3D plot of SEC in Fig. (11) shows that  SEC is violated at small values of $\lambda$ but it  is valid for $\lambda >75$ with $\alpha=1$.
A 3D plot of SEC for different  $\lambda$ with $\alpha=2$ is shown in Fig. (12), it is found that  SEC is obeyed in the range $-8 \pi < \lambda <- 4 \pi$  but it violates for other values. Further analysis of SEC for different $\lambda$  with positive $\alpha $ plotted in Fig. (13),  show that  SEC is found to obey once again when $\lambda > 75$.
Thus it comes out that existence of wormholes in the modified gravity depends on the parameters of the coupling constants and we note two different  cases: (i) NEC satisfies, SEC violates for which we need exotic matters and  (ii) both NEC and  SEC are satisfied simultaneously signalling wormhole solutions with normal matter.

\section{Result and Discussion}

In the paper wormhole solutions are obtained in the modified   $f(R,T)=R+ \; \alpha R^{2}+\lambda  \;T$ gravity with static spherically symmetric  metric. The field equations  are highly non-linear and intractable in known forms, we found the energy density, radial and transverse pressure of the wormhole solutions for 
a given shape function. As the equation of state of matter for wormhole is not known, we study the energy conditions obeyed in the theoretical framework with  gravitational coupling parameters making use of shape functions.

\subsection{\bf For the shape function I :  $b(r)=r_{0}e^{1-\frac{r}{r_{0}}}$}

We note the following:\\

$\bullet$  For transverse and radial pressures the plot NEC in Figs. (1) and (2) show that NEC is violated near the throat but away from the throat it obey for $\alpha=1$ with $\lambda >0$. For a given $\lambda$,  NEC is checked with different $\alpha$  in Figs. (2) and (3) which show that  NEC is satisfied for $\alpha >0$.  It is evident from Fig. (4) that the energy density  is   positive for $\alpha >0$.  For $\alpha < 0$ and $\lambda=0$ one obtains wormhole solutions as found in Ref. (\cite{35}).   In this case  WH exists in the presence of exotic matter.\\

$\bullet$ The radial variation of the effective EoS parameter in Fig. (13) show that for $\alpha= 0$ and $-1$ with $\lambda =1$ the EoS parameter for the shape function I, is always negative. The EoS parameter for positive values of $\alpha$ with $\lambda =1$ show that  the effective state parameter is always negative which is however undefined at 
\begin{equation}
e^{\frac{r}{r_{0}}-1} \; r^{2}=\alpha \; \left( \frac{72\pi+11\lambda}{12\pi+2\lambda} \right).
\end{equation}
Far away from the throat EoS parameter decreases which attains $\omega \rightarrow -0.6$.  The effective EoS parameter remains negative at late times  indicating the presence of exotic matter accommodating an accelerating universe.
A detail numerical analysis is displayed in  Table-1 and Table-2.\\

\begin{figure}[H]
\begin{center}
\includegraphics[scale=0.55]{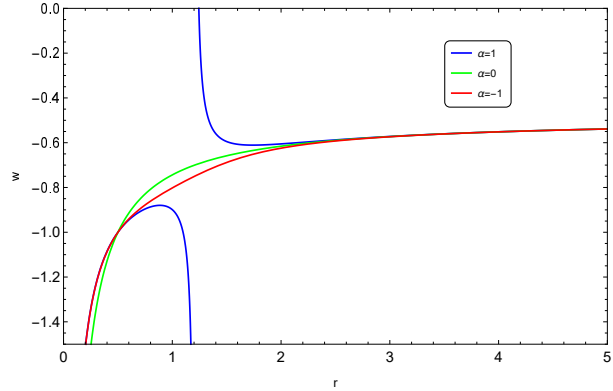}
\caption{EoS parameter, with different $\alpha$ and $\lambda=1$ for shape function I.}
\label{fig: 14}
\end{center}
\end{figure}

\begin{figure}[H]
\begin{center}
\includegraphics[scale=0.55]{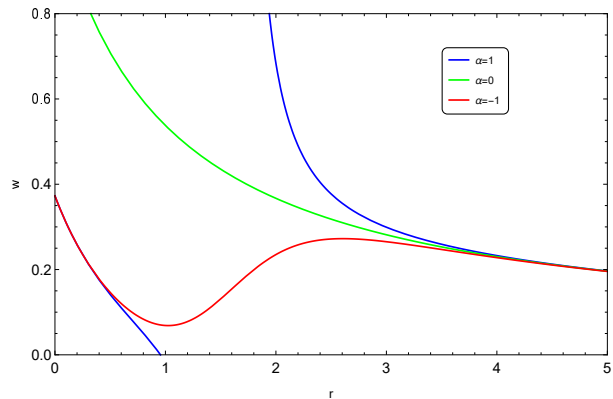}
\caption{EoS parameter, with different $\alpha$ and $\lambda=1$ for shape function II.}
\label{fig: 15}
\end{center}
\end{figure}

\subsection{\bf For the shape function : $b(r)=r_{0}^{2}\frac{e^{r_{0}-r}}{r}$ }
For a hybrid shape function we note the following\\
\\
$\bullet$  The energy density in Fig (5)  is found positive for $\lambda > - 4 \pi$. It is negative for the range   $-8 \pi<\lambda < - 4 \pi$ and for $\lambda < -8 \pi$  for $\alpha$ positive (say $\alpha=1$).

$\bullet$ Radial variation of energy density is  positive for $\alpha >0$ with $\lambda=1$ as shown in Fig. (6).

$\bullet$ Radial variation of NEC in Fig. (7) shows that it is valid from a small range of negative $\lambda$ to large positive values, thus we note existence of wormhole over a wide range of values of $\lambda$ which is not found in Ref. (\cite{45}).

$\bullet$ The radial variation of  energy conditions for  $\alpha =1$ and $\lambda=5$ shown in Fig. (8) shows that the energy conditions except SEC are valid. But for  $\alpha =1$ and large $\lambda$ ($\lambda=80$) in Fig. (9) it is evident that all the energy conditions are obeyed right from the wormhole throat. The later case permits wormhole solution with normal matter at the throat in $f(R, T)$-gravity. 

$\bullet$ SEC  plotted in Fig. (10) shows that it is valid for large $\lambda > 75.3$ for $\alpha$ positive. 
In Fig. (11) SEC is found to valid for a limited range of negative for $\alpha =2$.  In Fig. (12) we found that SEC is valid for $\lambda=80$ with positive $\alpha$ indicating that wormhole solutions  exist in the presence of non-exotic matter which however is not permitted in GR. We note that for  $\alpha <0$ all the energy conditions are not valid. \\

$\bullet$ The radial variation of the effective EoS parameter for different $\alpha$  is plotted in Fig.(14). The effective EoS is found always positive with $\lambda =1$. There is a minimum for $\alpha <0$ then increases attains a maximum thereafter it decreases  slowly. For  $\alpha> 0$ it is evident that $\omega$ decreases from a positive to a negative value  with a discontinuity  at 
\begin{equation}
\frac{2e^{r-r_{0}}r^{4}}{(1+r)r_{0}^{2}}=\alpha \; \left(\frac{ 72\pi+11\lambda}{6\pi+\lambda} \right).
\end{equation}
A detail analysis of the models for obtaining wormholes for shape functions II is displayed in Tables - (3) and (4).\\

$\bullet$ For $\lambda=0$, the modified gravity reduces to  the $f(R)$-gravity  and existence of wormhole  is studied considering with $n \neq 2$ in $f(R)=R+\alpha R^{n} $ theory of gravity for a given shape function  \cite{64}. In the present paper we consider  $n=2$ and different energy conditions are examined for obtaining wormhole solutions. It is found that the wormhole solutions  with exotic and non-exotic matter  are permitted for different values of the parameters $\alpha$ and $\lambda$. The result obtained here is different from  that derived in GR where energy conditions are violated indicating requirement of exotic matter. This is an interesting result which leads to a distinction between the GR and the modified theories of gravity.

\begin{table*}[t]
 \centering
  \begin{tabular}{|c|c|c|c|}\hline
Terms & $\lambda=1$ & $\lambda=-1$ & $\lambda=-30$  \\ \hline
$\rho$ & $\ge 0$ for $r\;\epsilon\; (0,\infty)$ & $\ge 0$ for  $r\;\epsilon\; (0,\infty)$ & $\le 0$ for $r\;\epsilon \;(0,\infty)$ \\ \hline

$\rho+p_{r}$ & $\ge 0$ for $r\;\epsilon\; (0,\infty)$  & $\ge 0$ for  $r\;\epsilon\; (0,\infty)$  & $\le 0$ for $r\;\epsilon\; (0,\infty)$ \\ \hline

$\rho+p_{t}$ & $<0$ for $r\;\epsilon\; (0,0.5)$ & $<0$ for $r\;\epsilon\; (0,0.5)$ & $<0$ for $r\;\epsilon\; (0,1.28)$ \\
&$>0$ for $r\;\epsilon\; (0.5,1.25)$ & $>0$ for $r\;\epsilon\; (0.5,1.3)$ & $\ge 0$ for $r\;\epsilon\; (1.28,\infty)$  \\
&$\le 0$ for $r\;\epsilon\; (1.25,\infty)$ & $\le 0$ for $r\;\epsilon\; (1.3,\infty)$ & \\ \hline

$\rho-|p_{r}|$ & $<0$ for $r\;\epsilon\; (0,0.25)$ & $<0$ for $r\;\epsilon\; (0,0.23)$ & $\le 0$ for $r\;\epsilon\; (0,\infty)$ \\
&$>0$ for $r\;\epsilon\; (0.25,1.3)$ & $>0$ for $r\;\epsilon\; (0.23,1.27)$ &   \\
&$\le 0$ for $r\;\epsilon\; (1.3,\infty)$ & $\le 0$ for $r\;\epsilon\; (1.27,\infty)$ & \\ \hline

$\rho-|p_{t}|$ & $<0$ for $r\;\epsilon\; (0,0.5)$ & $<0$ for $r\;\epsilon\; (0,0.5)$ & $\le 0$ for $r\;\epsilon\; (0,\infty)$ \\
&$>0$ for $r\;\epsilon\; (0.5,1.3)$ & $>0$ for $r\;\epsilon\; (0.5,1.3)$ &  \\
&$\le 0$ for $r\;\epsilon\; (1.3,\infty)$ & $\le 0$ for $r\;\epsilon\; (1.3,\infty)$ & \\ \hline

\end{tabular}
\caption{Summary of results for $b(r)=r_{0}e^{1-\frac{r}{r_{0}}}$ with $\alpha=2$ and different $\lambda$.}

 \label{tab:1}
\end{table*}


\begin{table*}[t]
 \begin{center}
  \begin{tabular}{|c|c|c|}\hline
Terms & $\alpha>0$ & $\alpha<0$ \\ \hline
$\rho$ & $\ge 0$ for $r\;\epsilon\; (0,\infty)$ & $\le 0$ for  $r\;\epsilon\; (0,\infty)$ \\ \hline

$\rho+p_{r}$ & $\ge 0$ for $r\;\epsilon\; (0,\infty)$ & $\le 0$ for  $r\;\epsilon\; (0,\infty)$ \\ \hline

$\rho+p_{t}$ & $<0$ for $r\;\epsilon\; (0,0.5)$ & $\le0$ for $r\;\epsilon\; (0,\infty)$ \\
&$>0$ for $r\;\epsilon\; (0.5,1.25)$ &   \\
&$\le 0$ for $r\;\epsilon\; (1.25,\infty)$ & \\ \hline

$\rho-|p_{r}|$ & $<0$ for $r\;\epsilon\; (0,0.25)$ & $\le0$ for $r\;\epsilon\; (0,\infty)$ \\
&$>0$ for $r\;\epsilon\; (0.25,1.3)$ &   \\
&$<\le 0$ for $r\;\epsilon\; (1.3,\infty)$ & \\ \hline

$\rho-|p_{t}|$ & $<0$ for $r\;\epsilon\; (0,0.5)$ & $\le0$ for $r\;\epsilon\; (0,\infty)$ \\
&$>0$ for $r\;\epsilon\; (0.5,1.3)$ &   \\
&$\le 0$ for $r\;\epsilon\; (1.3,\infty)$ & \\ \hline

\end{tabular}
\caption{Summary of results for $b(r)=r_{0}e^{1-\frac{r}{r_{0}}}$ with $\lambda=1$ and different $\alpha$.}
\end{center}
 \label{tab:2}
\end{table*}

\begin{table*}[t]
 \centering
  \begin{tabular}{|c|c|c|c|}\hline
Terms & $\lambda=1$ & $\lambda=76$ & $\lambda=-13$  \\ \hline
$\rho$ & $\ge 0$ for $r\;\epsilon\; (0,\infty)$ & $\ge 0$ for  $r\;\epsilon\; (0,\infty)$ & $\le 0$ for $r\;\epsilon\; (0,\infty)$ \\ \hline

$\rho+p_{r}$ & $\ge 0$ for $r\;\epsilon\; (0,\infty)$ & $\ge 0$ for  $r\;\epsilon\; (0,\infty)$ & $\ge 0$ for $r\;\epsilon\; (0,\infty)$ \\ \hline

$\rho+p_{t}$ & $\ge 0$ for $r\;\epsilon\; (0,\infty)$ & $\ge0$ for $r\;\epsilon\; (0,\infty)$ & Violated for a range of $r$ \\ \hline

$\rho-|p_{r}|$ & $\ge 0$ for $r\;\epsilon\; (0,\infty)$ & Violated for small $r$ & $\le0$ for $r\;\epsilon\; (0,\infty)$ \\ \hline

$\rho-|p_{t}|$ & $\ge 0$ for $r\;\epsilon\; (0,\infty)$ & $\ge 0$ for $r\;\epsilon\; (0,\infty)$ & $\le 0$ for $r\;\epsilon\; (0,\infty)$ \\ \hline

$\rho+p_{r}+2 p_{t}$ & $\le 0$ for $r\;\epsilon\; (0,\infty)$ & $\le 0$ for $r\;\epsilon\; (0,\infty)$ & $\ge 0$ for $r\;\epsilon\; (0,\infty)$ \\ \hline
\end{tabular}
\caption{Summary of results for $b(r)=r_{0}^{2}\frac{e^{r_{0}-r}}{r}$ with $\alpha=1$ and different $\lambda$.}

 \label{tab:3}
\end{table*}


\begin{table*}[t]
 \begin{center}
  \begin{tabular}{|c|c|c|}\hline
Terms & $\alpha>0$ & $\alpha<0$ \\ \hline
$\rho$ & $\ge 0$ for $r\;\epsilon\; (0,\infty)$ & $\le 0$ for  $r\;\epsilon\; (0,\infty)$ \\ \hline

$\rho+p_{r}$ & $\ge 0$ for $r\;\epsilon\; (0,\infty)$ & $\le 0$ for  $r\;\epsilon\; (0,\infty)$ \\ \hline

$\rho+p_{t}$ & $\ge 0$ for $r\;\epsilon\; (0,\infty)$ & $\le0$ for $r\;\epsilon\; (0,\infty)$ \\ \hline

$\rho-|p_{r}|$ & $\ge 0$ for $r\;\epsilon\; (0,\infty)$ & $\le0$ for $r\;\epsilon\; (0,\infty)$ \\ \hline

$\rho-|p_{t}|$ & $\ge 0$ for $r\;\epsilon\; (0,\infty)$ & $\le0$ for $r\;\epsilon\; (0,\infty)$ \\ \hline

$\rho+ p_{r}+2 p_{t}$ & $\le0$ for $r\;\epsilon\; (0,\infty)$ & $\ge0$ for $r\;\epsilon\; (0,\infty)$ \\ \hline
\end{tabular}
\caption{Summary of results for $b(r)=r_{0}^{2}\frac{e^{r_{0}-r}}{r}$ with $\lambda=1$ and different $\alpha$.}
\end{center}
 \label{tab:4}
\end{table*}

\begin{figure}[H]
\begin{center}
\includegraphics{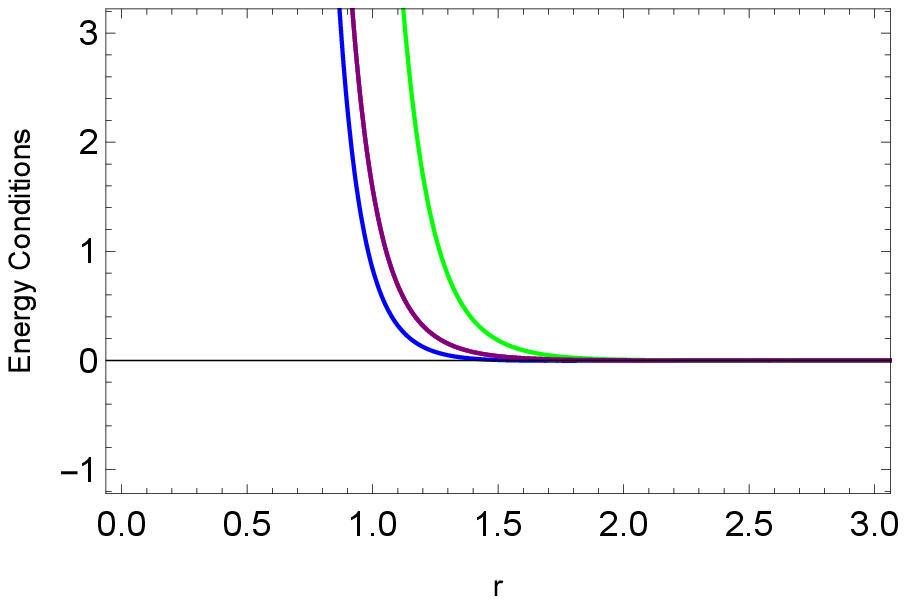}
\caption{NEC and DEC, with $\alpha=10$ and $\lambda=0$ for shape function II. Here $\rho+p_{r}>0$ (Blue), $\rho>|p_{r}|$ (Green) and $\rho>|p_{t}|$ (Purple).}
\label{fig: 16}
\end{center}
\end{figure}

$\bullet$  In the $f(R, T)$- gravity, wormhole solutions obtained with the shape function in Ref. (\cite{45}) is different from that considered in this paper. We note that there exists a class of wormholes with normal matter in our model with a wide range of values of  positive as well as negative values of $\lambda$. Thus  the choice of shape function is important as in Ref. (\cite{45}), wormholes are obtained  for negative values of $\lambda$ 
where NEC  for radial pressure is obeyed but the other two energy conditions namely,  NEC for transverse pressure and DEC for radial pressure are not obeyed, for positive $\lambda$, all the energy conditions are violated, a different result from that is obtained in  our model with shape function II. We note that in the limit  $\lambda \rightarrow 0$ and the plots of NEC and DEC with $\lambda=0$ is shown in Fig. (15) for shape function II, they are obeyed. Thus wormholes are obtained with normal matter in $f(R)$ model. We also note that  energy conditions are valid   in the  case of $f(R)=R+\alpha R^{2}$ with $\alpha \geq0$.
 The absence of observational signature of wormholes made  the geometrical and the material properties of these objects unpredictable, therefore,  wormholes are assigned  with a shape function and   EoS.  
 It is noted  that there exists a range of values of the coupling parameters which admits wormhole solution without  exotic matter in the case of second shape function which however is not possible in the case of the shape function I.  It is generally believed that the  present universe emerged out from an inflationary phase at an early era thereafter it enters into present accelerating phase followed by matter domination.   Another interesting scenario called emergent universe  \cite{56,57} where the universe  begins from a static Einstein Universe which transits to other phases and encompasses finally the present observed universe  may be realized in the framework of  wormholes geometry.  The wormhole throat may be considered as the seed of the early static Einstein phase and away from the throat the late accelerating phase may be realized which will be discussed elsewhere.

$\bullet$ The no go theorem in the modified $f(R,T)=R+\alpha R^2 + \lambda T$-gravity with isotropic pressure does not arise for $\lambda \neq - 8 \pi$ for the wormhole solutions unlike that obtains in GR \cite{26a}. However, the no go theorem for wormholes  in GTR \cite{26a} is revisited in $f(R,T)$-theory even with an anisotropic fluid when $\lambda = - 8 \pi$ which follows from eq. (20). This is a new observation in modified gravity considered here.
 Thus we explore existence of wormholes with anisotropic pressure of the fluid  for 
$\lambda \neq - 8 \pi$.
The radial variation of SEC for $\lambda=80$ for different $\alpha$  shown in fig (12) is interesting, it predicts that  for positive $\alpha$ all the energy conditions except the SEC is obeyed. However, when $\lambda$ is increased, it is found that for higher values of $\lambda$, say, $\lambda=80$, all the energy conditions including SEC are valid once again at the throat but  $\alpha \leq 0$.

\section{Acknowledgment}
AC would like to thank University of North Bengal for awarding Senior Research Fellowship. SD is thankful to UGC, New Delhi for financial support. The authors would like to thank IUCAA Centre for Astronomy Research and Development (ICARD), NBU for extending research facilities. The authors would like to thank anonymous referee for presenting the paper in its current form. BCP would like to thank DST-SERB Govt. of India (File No.:EMR/2016/005734) for a project.\\

 \pagebreak

\end{document}